\documentclass[aps,pre,showpacs,preprintnumbers,nofootinbib,floatfix,floats,groupedaddress,column]{revtex4}

\usepackage{bm}
\usepackage{latexsym}
\usepackage{dcolumn}
\usepackage{amsmath,amsfonts,amssymb}
\usepackage{graphicx,epsfig}
\usepackage{color}
\usepackage{verbatim}

\newcommand{\nab}{\mbox{\boldmath $\nabla$} {}}
\newcommand{\bra}[1]{\langle #1\rangle}
\newcommand{\meanemf}{\overline{\cal E} {}}
{}
\newcommand{\meanBB}{\overline{\mbox{\boldmath $B$}}{}}{}
\newcommand{\meanB}{\overline{\mbox{$B$}}{}}{}
\newcommand{\meanUU}{\overline{\mbox{\boldmath $U$}}{}}{}

\newcommand{\uu}{\mbox{\boldmath $u$} {}}

\newcommand{\bb}{\mbox{\boldmath $b$} {}}

\newcommand{\ex}{\mbox{{\boldmath $e$}}_{1}}
\newcommand{\ey}{\mbox{{\boldmath $e$}}_{2}}
\newcommand{\ez}{\mbox{{\boldmath $e$}}_{3}}
                                            
\newcommand{\beq}{\begin{equation}}
\newcommand{\eeq}{\end{equation}}

\newcommand{\bfa}{\textbf{a}}
\newcommand{\bfc}{\textbf{c}}
\newcommand{\bff}{\textbf{f}}
\newcommand{\bfq}{\textbf{q}}
\newcommand{\bfv}{\textbf{v}}
\newcommand{\bfx}{\textbf{x}}
\newcommand{\bfW}{\textbf{W}}

\newcommand{\bfE}{\textbf{E}}
\newcommand{\bfF}{\textbf{F}}
\newcommand{\bfG}{\textbf{G}}
\newcommand{\bfQ}{\textbf{Q}}

\newcommand{\tdamp}{\mbox{$\tau_{\rm d}$}}

\reversemarginpar

\begin{document}

\title{Mean field dynamo action in renovating shearing flows}
\author{Sanved Kolekar}
\email{sanved@iucaa.ernet.in}
\author{Kandaswamy Subramanian}
\email{kandu@iucaa.ernet.in}
\affiliation{IUCAA, Pune University Campus, Ganeshkhind,
 Pune 411007, INDIA.} 
\author{S.~Sridhar}
\email{ssridhar@rri.res.in}
\affiliation{Raman Research Institute, Sadashivanagar, 
Bangalore 560080, INDIA.} 

\date{\today}
\begin{abstract}

We study mean field dynamo action in renovating flows 
with finite and non zero correlation time ($\tau$) in the
presence of shear. Previous results
obtained when shear was absent are generalized to the case with shear.
The question of whether the mean magnetic field can grow
in the presence of shear and non helical turbulence, 
as seen in numerical simulations, is examined. We show in a  
general manner that, if the motions are strictly non helical,
then such mean field dynamo action is not
possible. This result is not limited to low (fluid or magnetic) 
Reynolds numbers nor
does it use any closure approximation; it only assumes 
that the flow renovates itself after each
time interval $\tau$. Specifying to a particular form
of the renovating flow with helicity, we recover the standard
dispersion relation of the $\alpha^2\Omega$ dynamo,
in the small $\tau$ or large wavelength
limit. Thus mean fields grow even in the
presence of rapidly growing fluctuations, surprisingly, in a manner
predicted by the standard quasilinear closure, even though such a 
closure is not strictly justified.
Our work also suggests the possibility
of obtaining mean field dynamo growth in the presence of
helicity fluctuations, although having a coherent helicity
will be more efficient. 

\end{abstract}


\maketitle
\vskip 0.5 in
\noindent
\maketitle

\section{Introduction} \label{sec:intro}

Understanding the origin of coherent large scale magnetic fields, 
observed in astrophysical systems from stars to galaxies, is of fundamental importance in astrophysics. The standard paradigm invokes the 
amplification of small seed magnetic fields due to dynamo action,
typically involving large scale shear flows combined with helical
turbulence \cite{BS05,dynam}. Recent numerical simulations have also 
raised the possibility that coherent fields can arise in non helical 
turbulent flows in the presence of shear \cite{shear_sim}, although the mechanism of how this happens is unclear \cite{shear_dyn,SS09}.

The dynamics of large--scale magnetic fields is generally described
by using the equations of mean field electrodynamics. Here one defines the
mean magnetic field $\meanBB$ and mean velocity field $\meanUU$
by suitable averaging over the small scales corresponding to the turbulent
fluctuations. The mean field evolution is
governed by the averaged version of the induction equation, which results 
in an extra term, the turbulent electromotive force 
$\meanemf = \bra{\uu \times \bb}$, where $\uu$ and $\bb$ are the fluctuating
velocity and magnetic fields.
Expressing $\meanemf$ in terms of the mean fields themselves is 
a closure problem, even for prescribed velocity fields. 
If the fluctuations can be assumed to be small,
one can employ the quasi--linear approximation or what is traditionally
known as the first order smoothing approximation (FOSA) to express 
the turbulent electromotive force $\meanemf$ 
in terms of a term proportional to $\meanBB$
(the $\alpha$--effect) and one proportional to the mean 
current density (turbulent diffusion).
The $\alpha$--effect, which depends on the helicity of turbulence is crucial
to amplify mean fields.

However, turbulent motions above some modest magnetic Reynolds number 
lead to a fluctuation dynamo and a rapid growth of 
magnetic noise \cite{ZRS83,Kaz68,BS05}. 
The growth rate of the fluctuation dynamo is 
typically larger than the growth rates associated
with the mean--field dynamo. In the presence of this 
rapidly growing magnetic noise, the validity of
the quasi--linear approximation or FOSA becomes suspect.
The question then arises whether one can indeed make sense of 
mean field concepts like the alpha effect.

In this context, considering exactly solvable flow models becomes 
very useful. For example, if one assumes a random flow whose correlation 
time is exactly zero, dynamo action, on both small and large scales can be studied in great analytical detail \cite{ZRS83,Kaz68,BS05}; 
however, such a flow is unphysical. Renovating flows discussed by several authors \cite{Ditt,GB92} (and references therein) provide, on the other hand, models involving flows with a finite non zero correlation time but which are 
still analytically tractable. 
In renovating flows time is split into successive intervals of
length $\tau$ and the stochastic component of the velocity in the different
intervals are assumed to be statistically independent realizations of an
underlying probability distribution (PDF). As the flow loses memory between different
time intervals, the evolution of the moments of the
magnetic field over any one time interval can be calculated by averaging
over the underlying PDF. Considering random helical renovating flows,
Gilbert and Bayly (GB) \cite{GB92} showed that the magnetic field 
becomes increasingly
intermittent with time. Nevertheless, the mean magnetic field can still grow
with a growth rate which approaches that of a standard mean--field $\alpha^2$ dynamo in the limit of small $\tau$ \cite{GB92,Ditt}. These works thus 
provide explicit demonstration that the growth of magnetic noise need not destroy the growth of the mean field even in the case of flows (which have 
this periodic loss of memory) with finite non zero correlation times.

In the present work we generalize some of these results to renovating flows
incorporating also a large scale shear. Our primary motivation is 
to examine if the introduction of shear can lead to a mean field growth even if the stochastic velocity field is non helical. 
It turns out that, once properly formulated in 
terms of renovating shearing waves, the details of our calculation share some crucial features with those of Gilbert and Bayly (GB)~\cite{GB92}. However,  GB have not given these steps; so to aid the presentation of our results, we begin in \S~2 with a presentation of the main results of \cite{GB92} for the mean field evolution in renovating helical flows. In \S~3 we formulate the problem of renovating flows with a background linear shear, and prove a general result
that there is no dynamo action when  the flow is strictly non helical.
We also consider a particular example of renovating shearing waves with 
helicity, caused by overdamped external forcing and recover the dispersion relation for the $\alpha^2\Omega$ dynamo. The final section summarizes and presents a discussion of our results. 

\section{Renovating helical flows without shear}

We re--derive here the results of Gilbert and Bayly~\cite{GB92} on mean field evolution in a model helical renovating flow, in the absence of background shear. Consider the induction equation for the evolution of the magnetic field,
\begin{equation}
\frac{\partial \textbf{B}}{\partial t} \;=\; 
\nab \times \left[ \textbf{u} \times \textbf{B} - 
\eta \nab \times \textbf{B} \right].
\label{inductioneqn}
\end{equation}
GB assumed the velocity field $\textbf{u}$
to have zero mean 
with only a turbulent component. In each renovating time interval $\tau$, they took
\begin{equation}
\textbf{u}({\bf x})={\bf a}\sin({\bf q}\cdot {\bf x}+\psi)+{\bf b}h \cos({\bf q}\cdot {\bf x}+\psi)
\label{uturbdef}.
\end{equation}
with the conditions
\begin{equation}
{\bf a}\cdot {\bf q}=0, \quad  {\bf b}={\bf q} \times {\bf a}/q
\end{equation}
which implies incompressibility, that is $\nabla \cdot \textbf{u} =0 $.
The parameter $h$ satisfies $-1 \leq h \leq 1$ and determines the helicity of the flow. This helical flow is made random by choosing the parameters of the flow randomly and independently from an underlying PDF, for every time interval $\tau$. The ensemble considered is the following: In each time interval $\left[(n-1)\tau, n\tau\right]$, (i) $\psi$ is chosen uniformly random between 0 to $2\pi\,$; (ii) the propagation vector {\bf q} is uniformly distributed on a sphere of radius $q\,$; (iii) for every fixed ${\bf q}$, the direction of 
${\bf a}$ is uniformly distributed in a circle of radius $a$ in the plane perpendicular to {\bf q}. The parameters $(q,a,h,\tau)$ are non--random and completely describe the renovating flow. The randomness of $\psi$ in condition (i) ensures statistical homogeneity, whereas conditions (ii) and (iii) ensure statistical isotropy of the flow.

The evolution of the magnetic field from time $(n-1)\tau$ to $n\tau$
is given by 
\begin{equation}
 B_i({\bf x},n\tau) \;=\; \int \mathcal G_{ij}({\bf x},{\bf y}) 
B_j( {\bf y},(n-1)\tau) \ d^3y
\label{evolutioneqn}
\end{equation}
where $\mathcal G_{ij}({\bf x},{\bf y})$ is the Green's function of the induction equation Eq.(\ref{inductioneqn}). $\mathcal G_{ij}$ is random due 
to the randomness of the turbulent velocity field. We take the ensemble 
average of this equation over the ensemble described above and note that the
velocity in any time interval $\left[(n-1)\tau, n\tau\right]$ is uncorrelated with the initial magnetic field at time $(n-1)\tau$. Thus the average of the product of $\mathcal G_{ij}$ and $B_j$ can be written as the product of the averages, an important simplification arising from the loss of memory of renovating flows. The mean field $\meanBB$ then evolves as,
\begin{equation}
\meanB_i({\bf x},n\tau) \;=\; \int \overline{\mathcal G_{ij}}({\bf x},{\bf y}) 
\meanB_j( {\bf y},(n-1)\tau) \ d^3y .
\label{evolutioneqn2}
\end{equation}
Further, from the statistical homogeneity of the renovating flow, one has
$\overline{\mathcal G_{ij}}({\bf x},{\bf y}) = \overline{\mathcal G_{ij}}({\bf x}-{\bf y})$; then Eq.~(\ref{evolutioneqn2}), which is a convolution in physical space, becomes a product in Fourier space. Defining the spatial Fourier transform, 
\begin{equation}
\hat{\meanB}_i({\bf k}, t) \;=\; \int d^3x \ \meanB_i({\bf x},t) e^{-i{\bf k}\cdot{\bf x}}
\label{four}
\end{equation}
we have in Fourier space,
\begin{equation}
\hat{\meanB}_i({\bf k},n\tau) \;=\; G_{ij}({\bf k}) \times \hat{\meanB}_j({\bf k},(n-1)\tau)
\label{fourevol}
\end{equation}
where the response tensor, $G_{ij}({\bf k})$, is defined by,
\begin{equation}
G_{ij}({\bf k}) \;=\; \int  \overline{{\mathcal G}_{ij}}({\bf x}-{\bf y}) 
e^{-i{\bf k}\cdot ({\bf x}-{\bf y})} \ d^3x  .
\label{responsetensor}
\end{equation}
The mean magnetic field will grow exponentially if its Fourier component is a 
eigenvector of the matrix $G_{ij}({\bf k})$ with eigenvalue $\sigma$, whose magnitude is greater than one. 
For such an eigenvector we have
\begin{equation}
\hat{\meanB}_i(\textbf{k},n\tau) = \sigma^n\,\hat{\meanB}_i(\textbf{k},0) 
\quad {\rm or} 
\quad \hat{\meanB}_i(\textbf{k},t) = \sigma^{t/\tau}\, 
\hat{\meanB}_i(\textbf{k},0) 
\end{equation}
and so the growth rate of this eigenmode is given by
\begin{equation}
\lambda = \frac{\ln(\sigma)}{\tau}. 
\label{growthrate}
\end{equation}
Hence, the response tensor $G_{ij}({\bf k})$ contains all the information about 
the growth or decay of the mean magnetic field. We now proceed to calculate it for the renovating velocity field of Eq.~(\ref{uturbdef}).

\subsection{The response tensor} 

In order to explicitly calculate the response tensor for the evolution 
of the mean field in the renovating helical flow, an important simplification was introduced by GB. The renovation time  $\tau$ was split into two equal 
sub-intervals. In the first sub--interval (step~1) resistivity was neglected
and the field was just frozen--in and advected with the fluid, with twice the original velocity. In the second sub--interval (step~2), induction by the velocity was neglected and the field was assumed to diffuse with twice the resistivity. Although such an assumption seems plausible in the limit of a short renovation time---as this would be one way of numerically integrating the induction equation--- GB did not give any rigorous justification. For the present purpose, we adopt the same simplification as GB. Thus we consider the evolution of the magnetic field in these two steps and then Fourier transform the resulting averaged Green function.

\noindent
\textbf{Step~1}: During the time interval $0$ to $\tau /2$ we assume 
$\eta = 0$, and double the value of the velocity field. Then Eq.(\ref{inductioneqn}) becomes just the ideal induction equation,
\begin{equation}
\frac{d\textbf{B}}{dt} \;\equiv\; \left(\frac{\partial}{\partial t} + 2\textbf{u} \cdot \nabla\right)\textbf{B} \;=\; \left(\textbf{B}\cdot \nabla\right)2\textbf{u}
\end{equation}
with the standard Cauchy solution given by 
\begin{equation}
B_{i}({\bm r} ,t) \;=\; \frac{\partial r_{i}}{\partial y_{j}}
B_{j}({\bf y},t_{0})=J_{ij}({\bm r}) B_{j}({\bf y},t_{0}).
\end{equation}
Here $\textbf{B}({\bf y},t_{0})$ is the initial magnetic field; 
${\bm r}(t)$ is the position of a fluid element at time $t$, 
which was originally
at a `Lagrangian' position $\textbf{y}$ at time $t_0$. Note that the fluid elements follow
the integral curves of the velocity field, with 
\begin{equation}
d{\bf x}/dt = 2 {\bf u} \;=\; 2{\bf a}\sin(\Phi) +2{\bf b}h \cos(\Phi)
\label{traj}
\end{equation}
where we have substituted from Eq.~(\ref{uturbdef}), 
assumed twice the velocity for step 1 and defined the phase
$\Phi = {\bf q}\cdot {\bf x}+\psi$.
From the incompressibility condition, we have 
$d\Phi/dt = 2{\bf q}\cdot{\bf u} = 0$ and thus  
Eq.~(\ref{traj}) can be integrated to give at time $t=\tau/2$,
\begin{equation}
{\bm r} \;=\; {\bf y} + \tau \textbf{u} 
\;=\; {\bf y} + {\bf a} \tau \sin{\left( {\bf q} \cdot {\bf y} 
+\psi \right)} + {\bf b} \tau \text{h} \cos{\left( {\bf q}\cdot {\bf y} + 
\psi \right)}.
\end{equation}
Here we have used the constancy of the phase $\Phi$ and set it equal 
to its initial value $\Phi = {\bf q}\cdot {\bf y}+\psi$. Thus the Jacobian is
\begin{equation}
J_{ij}({\bm r}) \;\equiv\; \frac{\partial r_{i}}{\partial y_{j}} \;=\; 
\delta_{ij} + a_{i}q_{j}\tau \cos{\left({\bf q}\cdot {\bf y} + \psi \right)} - b_{i}q_{j}h\tau \sin{\left({\bf q}\cdot {\bf y} + \psi \right)}
\label{Jdef}
\end{equation}

\noindent
\textbf{Step~2}: During the time interval $\tau /2$ to $\tau $ the turbulent velocity field is zero 
and there is only diffusion present, with a resistivity $2\eta$. 
The induction equation then reduces to a 
diffusion equation for the magnetic field as
\begin{equation}
\frac{\partial \textbf{B}}{\partial t} \;=\; 2\eta {\nabla}^2 \textbf{B}
\end{equation}
The solution of this equation is given in terms of the resistive 
Green's function 
\begin{equation}
\mathcal G^{\eta}({\bf x}-{\bm r}) \;=\; \frac{1}{(4\pi\eta \tau)^{3/2}} \  
\exp{\left[-\,\frac{({\bf x}-{\bm r})^2}{4\eta \tau}\right]}
\end{equation}
The total Green's function defined in Eq.~(\ref{evolutioneqn}) 
is simply the product of the two Green's functions in the above two steps
\begin{eqnarray}
\mathcal G_{ij}({\bf x},{\bf y}) \;=\; \mathcal G^{\eta}({\bf x}-{\bm r}({\bf y})) J_{ij}({\bm r}({\bf y})),
\label{Gtotal}
\end{eqnarray}
where we have written explicitly ${\bf r}$ as a function of ${\bf y}$.
The response tensor defined in Eq.~(\ref{responsetensor}) then becomes
\begin{equation}
G_{ij}({\bf k}) \;=\; \overline{\int \frac{1}{(4\pi\eta \tau)^{3/2}} \  
e^{-\frac{({\bf x}-{\bm r})^2}{4\eta \tau}} \ 
J_{ij}({\bm r}) e^{-i{\bf k}\cdot ({\bf x}-{\bf y})} \ d^3x} 
\;=\; \overline{J_{ij} \ ({\bm r}({\bf y})) 
e^{-i{\bf k}\cdot ({\bm r}({\bf y})
- {\bf y})}}e^{-\eta \tau k^2},
\label{Gbeforeavg}
\end{equation}
where in the second step we have done the integral over ${\bf x}$.
The overhead bars in Eq.~(\ref{Gbeforeavg}) denote ensemble averages,
and we will see below that due to the statistical homogeneity of the
renovating flow, this averaged quantity does not depend on ${\bf y}$, but
only on ${\bf k}$.  Also note that in typical astrophysical systems, 
the resistive timescale will be much larger than the renovation time
and the value of $\eta k^2 \tau$ is typically much smaller than unity, 
and so can safely be set to zero. A non zero but small $\eta$ will decrease 
the growth rate by a negligible amount. 

\begin{figure}[!h]
\centerline{\includegraphics[width=6cm, height=8cm]{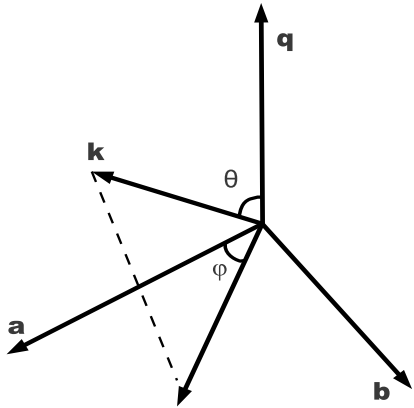}}
\caption{\label{axis} The different angles defined in the text. 
The angle between ${\bf k}$ and ${\bf q}$
is $\theta$.  The vectors ${\bf a}$, ${\bf b}$ and ${\bf q}$ are mutually
perpendicular to each other. The component of ${\bf k}$ perpendicular 
to ${\bf q}$ makes an angle $\phi$ with ${\bf a}$.} 
\end{figure}

We now evaluate the ensemble average in Eq.~(\ref{Gbeforeavg}). 
Note that GB state the final result, omitting all intermediate steps.
We give the detailed steps in Appendix~\ref{averaging} since they are of use in the case when shear is present. Here we list the important intermediate steps 
and the final expression. Let the angle between ${\bf k}$ and ${\bf q}$
be $\theta$; we will treat this as a colatitude and denote the azimuthal angle
of ${\bf q}$ by $\tilde{\phi}$. 
Let the component of ${\bf k}$ perpendicular to ${\bf q}$ make 
an angle $\phi$ with ${\bf a}$ (see Fig.~{\ref{axis}}). 
On averaging over phase $\psi$ we get
\begin{equation}
G_{ij}({\bf k}) \;=\; \overline{\delta_{ij}\ J_{0}(\tau ak \chi \sin\theta) 
- \frac{ihaq\tau}{\chi k\sin\theta} 
\left[\epsilon_{imn}k_{m}\hat{q}_{n}\hat{q}_{j} \right]J_{1}(\tau ak\chi 
\sin\theta)}
\label{Ghel}
\end{equation}
where $\chi  =  (\cos^2\phi + h^2 \sin^2 \phi)^{1/2}$ and the overhead bars 
denote ensemble averages over the remaining random variables $\theta$, $\tilde{\phi}$ and $\phi$. Averaging over direction of $\textbf{q}$ (i.e. averaging over the angles $\theta$ and $\tilde{\phi}$, we have
\begin{equation}
G_{ij}({\bf k}) \;=\; \delta_{ij}\ g_{0}(\tau ak, h) \;+\; \frac{ih\tau aq\epsilon_{ijm}k_{m}}{2k} g_{1}(\tau ak, h)\,,
\label{Gfinal}
\end{equation}
where 
\begin{equation}
g_{0}(s, h) \;=\; \overline{\frac{\sin(s\chi)}{s\chi}}\,;\qquad 
 g_{1}(s, h) \;=\; \overline{\frac{1}{\chi}\left( \frac{\sin(s\chi)}{(s\chi)^2}
-\frac{\cos(s\chi)}{s\chi} \right)}
\end{equation}
and now the overhead bars denote ensemble averages over the random variable 
$\phi$ (for maximally helical flow with $h = \pm 1$, so $\chi = 1$, and the response tensor becomes independent of the random variable $\phi$. In the rest
of this paper $h$ can take any value between $-1$ and $+1$). 
We recall that in step~1, we evolved the field for a time interval $\tau/2$, but with twice the velocity (i.e. $2a$ instead of $a$); nevertheless
the combination $a\tau$ which appears in the above equation remains
unaffected. 

One of the eigenvectors of $G_{ij}$ is $\textbf{k}$ with eigenvalue unity. 
But this eigenvector can be ignored since the magnetic field mode must be orthogonal to $\textbf{k}$ (i.e. we must have $\nab\cdot{\bf B}=0$). The relevant eigenvectors of $G_{ij}$ are found to be $(-i,1,0)^{\rm T}$ and $(i,1,0)^{\rm T}$, with the corresponding eigenvalues, $\sigma_{+}$ and $\sigma_{-}\,$, 
given by
\begin{equation}
\sigma_{\pm}=g_{0}(\tau ak, h) \;\mp\; \frac{\tau aqh}{2}g_{1}(\tau ak, h)
\end{equation} 
The growth rates $\lambda_{\pm}$, are given by
\begin{equation}
\lambda_{\pm} \;=\; \frac{1}{\tau}\ln\left(\sigma_{\pm}\right) \;=\; 
\frac{1}{\tau}\ln\left[g_{0}(\tau ak, h) \,\mp\, \frac{\tau aqh}{2}g_{1}(\tau ak, h)\right].
\end{equation}
Here we have divided by the full time interval $\tau$ to get the 
growth rate. Following GB it is readily verified that both $\pm$ mean 
field modes can grow for sufficiently large renovation times. GB also demonstrated 
that the magnetic field becomes increasingly intermittent, in the sense that higher order single--point moments of the field grow faster. Therefore, as advertised, the mean field dynamo operates efficiently in this case even in the presence of strongly growing magnetic noise.
It is of interest to look at the growth rate in the limit of
small renovation times such that $ak\tau \ll 1$ (this limit can also be looked upon as a long wavelength, or small $k$, limit at a fixed $\tau$), when 
\begin{eqnarray}
\lambda_\pm &\;\simeq\;& \mp\,k \frac{ha^2\tau q}{6} \;-\; k^2 \frac{(1 + h^2)a^2\tau }{12}\nonumber\\[2ex]
&\;\simeq\;& \pm\,\alpha k \;-\; \eta_{t} k^2\,,
\end{eqnarray}
where $\alpha = -(1/3)\overline{\textbf{u}\cdot (\nabla \times \textbf{u})}(\tau/2)$ and $\eta_t = (1/3)\overline{\textbf{u}\cdot \textbf{u}}(\tau/2)$ are the turbulent transport coefficients.\footnote{In these expressions for 
$\alpha$ and $\eta_t$, the factor $\tau/2$ appears instead of $\tau$, because the transport coefficients are really time integrals of correlation functions; 
see Eqs.~(6.19) and (6.20) in \cite{BS05}. For renovating flows in which the 
velocity field ${\bf u}$ acts over the full interval $\tau$ (i.e. if we were dealing with Eq.[inductioneqn], without the two--step prescription of GB), this implies averaging $t$ over the time interval $(0, \tau)$, which is equal to $\tau/2\,$. However, we get the same result even for the two--step prescription of GB, because two effects contribute in precisely opposite ways: when the velocity field is doubled in value, the transport coefficients quadruple because they are quadratic in the velocities; however, the doubled velocity field is ON for only the first half of any time interval $\tau$. Hence, 
averaging ``$t$'' over the time interval $(0, \tau)$ now implies integrating 
$t$ over the time interval $(0,\tau/2)$ and then dividing it by the full interval $\tau$, which gives $\tau/8$. Thus we obtain $4\times \tau/8 = \tau/2\,$.} Thus the growth rate for the case of small $ak\tau$ is identical to the growth rate of the standard $\alpha^2$ mean-field dynamo usually obtained using the quasilinear approximation or FOSA. We now turn to consider the influence of shear.

\section{Renovating shear flows} 

We now investigate the evolution of the mean magnetic field in scenarios 
when there is a mean shear flow over and above 
the background turbulence. 
Shear flows and turbulence are ubiquitous in astrophysical systems.
Recent work suggests that the presence of shear may open new
pathways to the operation of large--scale dynamos
\cite{shear_sim,shear_dyn,SS09}.
For simplicity we consider the background mean velocity
to be a linear shear flow.
Let $(\ex,\ey,\ez)$ be the unit vectors of a 
Cartesian coordinate system in the lab frame,
${\bf x} = (x_1,x_2,x_3)$ the position vector.
Without loss of generality, we choose this mean velocity to be in the
${\bm e_2}$ direction and varying linearly with $x_1$.
Thus the velocity field is given by 
\begin{equation}
\textbf{u}({\bf x},t) \;=\; {\bf u}_{\rm sh} \,+\, 
{\bf u}_{\rm turb} \;=\; Sx_1{\bm e_2}  \,+\, {\bf u}_{\rm turb}
\label{shearvelocity}
\end{equation}
where $S$ is the constant rate of shear coefficient. 
The turbulent velocity field ${\bf u}_{\rm turb}$ is composed of renovating shearing waves with quite general amplitudes at this point. In particular we take 
\begin{equation}
\textbf{u}_{\rm turb}({\bf x},t) \;=\; {\bf A}(t,{\bf q})
\sin\left[{\bf Q}(t)\cdot {\bf x}+\psi\right] \,+\,
{\bf C}(t,{\bf q}) \cos\left[{\bf Q}(t)\cdot {\bf x}+\psi\right],
\label{usheargen}
\end{equation}
where the wavevector is a shearing wavevector of
the form ${\bf Q} = (Q_1,Q_2,Q_3) \equiv (q_1-Sq_2(t-t_i), q_2, q_3)$,
and ${\bf q} = (q_1, q_2, q_3)$ its initial value at the beginning of
each renovation period, i.e at $t=t_i$. 
Note that ${\bf q}$ is chosen randomly from 
a specified PDF (see below) for each renovating period.
We will see that such a form of ${\bf u}_{\rm turb}$ naturally arises 
when we consider Fourier modes of the velocity which satisfy the momentum
equation in a background linear shear flow.
We will also later adopt explicit forms of ${\bf A}(t,{\bf q})$ and 
${\bf C}(t,{\bf q})$;
but several of the conclusions that we arrive at are quite general  
insensitive to the functional forms of ${\bf A}$ and ${\bf C}$. 
We also assume the turbulence to be 
incompressible with $\nabla \cdot \textbf{u}_{\rm turb} =0$, which implies 
\begin{equation}
 \textbf{Q}(t) \cdot \textbf{A}(t) \;=\; 0 \,; \qquad
\textbf{Q}(t) \cdot \textbf{C}(t) \;=\; 0
\end{equation}
Thus the amplitudes have to shear in an opposite sense to the wavevector so
as to maintain incompressibility.
The shearing wavevector can be written in a compact form as
$Q_j = q_i\gamma_{ij}(-(t-t_i))$, where 
$\gamma_{ij}(t)$ is the shearing matrix defined by, 
\begin{equation}
\gamma_{ij}(t) \;=\; \delta_{ij} \,+\, \delta_{i2}\delta_{j1} St.
\end{equation}
The helicity of the turbulent velocity field $\textbf{u}_{\rm turb}({\bf x},t)$ is 
\begin{equation}
H = \textbf{u}_{\rm turb} \cdot (\nabla \times \textbf{u}_{\rm turb}) =
 \textbf{C} \cdot (\textbf{Q} \times \textbf{A}),
\label{helicityshear}
\end{equation}
and this vanishes unless both $\textbf{C}$ or $\textbf{A}$ are nonzero.

As in the previous section we consider the turbulent flow 
to be a pulsed renovating flow. 
The turbulent velocity field is assumed to be ON for a 
time interval $\tau /2$, with twice its amplitude and
with diffusion absent. For the next $\tau /2$ interval, 
the turbulent velocity 
field is OFF and only the diffusion is present 
with resistivity $2\eta$. 
On an average, the turbulent velocity field is then correlated only for a 
time interval $\tau$. The mean shear flow on the other hand 
is always present for the full time interval $\tau$. 
The turbulent flow is randomized 
by considering an ensemble similar to that assumed
for the renovating flow without shear.
In each time interval $\left[(n-1)\tau, n\tau\right]$ 
(i) $\psi$ is chosen uniformly random between 0 to $2\pi$ 
(ii) the propagation vector {\bf q} is uniformly distributed 
on a sphere of radius $q$.
The randomness of $\textbf{A}$ is decided by the explicit 
form of $\textbf{A}$ itself, which we fix later, when solving
for the explicit form of the response tensor. 
For the general analysis we will not require it. 
The response tensor will be essential here too to determine 
the growth or decay of the mean magnetic field modes.

\subsection{The response tensor} 

We compute the response tensor for the evolution of the mean magnetic 
field again in two steps. 

\noindent
\textbf{Step~1}: During the time interval $t=0$ to $t=\tau/2$, 
$\eta = 0$ and Eq.~(\ref{inductioneqn}) reduces
again to the ideal induction equation, whose solution is as before
the Cauchy solution
\begin{equation}
B_{i}({\tilde {\bm r}} ,t) \;=\; \frac{\partial {\tilde r}_{i}}{\partial y_{j}}B_{j}({\bf y}, 0) \;=\; J_{ij}({\tilde {\bm r}}) B_{j}({\bf y},0)
\end{equation}
Here ${\tilde {\bm r}}$ gives the location of the fluid element at time
$t$, which was at time $t=0$ at the location $\textbf{y}$. These
positions are now on the integral curve of the sheared and turbulent velocity field, and so the trajectory now
obeys
\begin{equation}
d{\bf x}/dt \;=\; {\bf u}_{\rm sh} \,+\, 2{\bf u}_{\rm turb} \;=\;
Sx_1\ey \,+\, 2{\bf A}(t,{\bf q})
\sin{\tilde\Phi} \,+\, 2{\bf C}(t,{\bf q}) \cos{\tilde\Phi}.
\label{trajshear}
\end{equation}
We have substituted here from Eq.~(\ref{shearvelocity}),
assumed twice the turbulent velocity for step~1 and defined the phase
${\tilde\Phi} = [{\bf Q}(t)\cdot {\bf x}+\psi]$.
Note that we have not doubled the shear velocity, as we keep the shear
flow throughout the full period $(0,\tau)$.
From the incompressibility condition, we have ${\bf Q}\cdot{\bf u}_{\rm turb} =0$.
Therefore 
\begin{equation}
d{\tilde\Phi}/dt \;=\; {\dot{\bf Q}}\cdot {\bf x} 
\,+\, {\bf Q}(t)\cdot {\dot{\bf x}} \;=\;  
-Sq_2x_1 \,+\, Sq_2x_1 \,+\, 2{\bf Q}\cdot{\bf u}_{\rm turb} \;=\; 0.
\label{incom}
\end{equation}
The constancy of ${\tilde\Phi}$ can be used to express it in terms
of the initial position of the fluid element ${\bf y}$, and the initial
wavevector ${\bf Q}(t=0) = {\bf q}$, that is we can write ${\tilde\Phi}
= \textbf{q} \cdot \textbf{y} + \psi$. Then Eq.~(\ref{trajshear}) can be integrated to give
\begin{eqnarray}
{\tilde r}_i &\;=\;& \gamma_{ij}(t)r_j\nonumber\\[2ex]
r_j &\;=\;& {\tilde a}_j(t,q_k)
\sin{\left( \textbf{q} \cdot \textbf{y} + \psi \right)} 
+ {\tilde c}_j(t,q_k) \cos{\left( \textbf{q} \cdot \textbf{y} + \psi \right)} 
+ y_j 
\label{integralcurveshear}
\end{eqnarray}
where $r_j$ is a sheared position vector that will be of use later, 
and the coefficients ${\tilde a}_j$ and ${\tilde c}_j$ are defined by, 
\begin{eqnarray}
{\tilde a}_j(t,q_i) &\;=\;& \gamma_{jp}(-t) \int_0^t 2{A}_p(t^{\prime},q_i) 
dt^{\prime} \;+\; S\delta_{j2} \int_0^t \int_0^{t^{\prime}} 
2A_1(t^{\prime \prime},q_i) dt^{\prime \prime} dt^{\prime}\nonumber\\[2ex]
{\tilde c}_j(t,q_i) &\;=\;& \gamma_{jp}(-t) \int_0^t 2{C}_p(t^{\prime},q_i) 
dt^{\prime} \;+\; S\delta_{j2} \int_0^t \int_0^{t^{\prime}}
2 C_1(t^{\prime \prime},q_i) dt^{\prime \prime} dt^{\prime}
\label{adefcdef}
\end{eqnarray}
Therefore the Jacobian matrix is, 
\begin{equation}
J_{ij}({\tilde {\bm r}}({\bm y}),t) \;=\; 
\frac{\partial {\tilde r}_{i}}{\partial y_{j}} \;=\;
\gamma_{ip}(t) \left [ \delta_{pj}
\,+\, {\tilde a}_{p}(t) q_{j} \cos{\left({\bf q}\cdot {\bf y} 
\,+\, \psi \right)} \,-\, {\tilde c}_{p}(t) q_{j} 
\sin{\left({\bf q}\cdot {\bf y} \,+\, \psi \right)} \right],
\label{Jdef2}
\end{equation}
We need this Jacobian to be evaluated at the time $t=\tau/2$. 

\noindent
\textbf{Step 2}: During $\tau /2$ to $\tau $ the turbulent 
velocity field is zero and there is diffusion present along with shear. 
The induction equation then reduces to the following form
\begin{equation}
\left( \frac{\partial}{\partial t} + \textbf{u}_{\rm sh} \cdot \nabla \right)\textbf{B}= ({\bf B}\cdot \nabla ){\bf u}_{\rm sh} + 2\eta {\nabla}^2 \textbf{B}
\end{equation}
The sheared Green's function for this equation is \cite{shearGF}
\begin{equation}
\mathcal G^{\eta}_{ij}\left({\bf x},{\tilde {\bm r}},\frac{\tau}{2}\right)
\;=\; \frac{\gamma_{ij}(\tau /2)}{(4 \pi \eta \tau)^{3/2}} 
\left( \text{det}|{\bf D}|\right)^{1/2}
\text{exp}\left(\frac{-{\tilde x}_i D_{ij}{\tilde x}_j}{4\eta \tau} \right) 
\end{equation}
where ${\tilde x}_i = \gamma_{ij}(-\tau/2) x_j-{\tilde r}_i$, and $D_{ij}$ 
is a symmetric matrix whose inverse is given by 
\begin{equation}
{\bf D^{-1}} \;=\; \left[ {\begin{array}{ccc}
 1 \;\;&\;\; -S\tau/4 \;\;&\;\; 0  \\[2ex]
 -S\tau/4 \;\;&\;\; 1 + (S\tau)^2/12 \;\;&\;\; 0 \\[2ex]
 0 \;\;&\;\; 0 \;\;&\;\; 1 \\
 \end{array} } \right]
\end{equation}

Note that in the above computation we have taken the initial time to be 
$t=0$ and the final time
as $t=\tau$; but the same steps and calculations go through for any renovating time interval
$\left[(n-1)\tau,n\tau\right]$, 
with the initial time being $t=(n-1)\tau$, the final time being $t=n\tau$, and the initial wavevector
${\bf q} = {\bf Q}((n-1)\tau)$. Hence during any such time interval $\tau$, the magnetic field at time $t = n\tau$ is related to the magnetic field at time $t = (n-1)\tau$ by
\begin{equation}
B_i({\bf x},n\tau) \;=\; \int \mathcal G^{\eta}_{ik}\left({\bf x},{\tilde {\bm r}},\frac{\tau}{2}\right) \ J_{kj}\left({\tilde {\bm r}}({\bm y}),\frac{\tau}{2}\right) \ B_j({\bf y},(n-1)\tau) \ d^3 {\tilde r}.
\label{shearB}
\end{equation}
We would like to calculate the response tensor starting from the above
evolution equation. For this we first
define the Fourier transform of ${\bf B}_i({\bf x},t)$ by expressing
it in terms of the shearing waves,
\begin{equation}
\hat{B}_i({\bf k},t) = \int B_{i}({\bf x},t) e^{-i{\bf K}(t) \cdot {\bf x}} d^3x 
\label{shearFour}
\end{equation}
where we have defined the sheared wavevector
$K_j(t) =k_i \gamma_{ij}(-(t-t_i)) \equiv (k_1-Sk_2(t-t_i), k_2, k_3)$
and ${\bf k}= {\bf K}(t_i)$ is the initial wavevector at time $t_i$,
which for each step we take to be the time $(n-1)\tau$.
We will see that the evolution of the mean magnetic field is 
especially transparent
when the field is expanded in terms of shearing waves in Fourier space. 

We take the Fourier transform of Eq.~(\ref{shearB}), and 
change the integration variable from ${\tilde {\bm r}}$ to ${\bf y}$.
The Jacobian for this transformation is unity as the flow velocity which
maps ${\bf y}$ to ${\tilde {\bm r}}$ is divergence free
i.e. $d^3{\tilde r} = d^3{y}$. We get 
\begin{eqnarray}
{\hat B}_i({\bf k}, n\tau) &\;=\;& 
\int B_{i}({\bf x},n\tau) e^{-i{\bf K}(n\tau) \cdot {\bf x}} \ d^3x 
\nonumber \\
&=& \int \int  \mathcal G^{\eta}_{ik}
\left({\bf x},{\tilde {\bm r}},\frac{\tau}{2}\right) \ 
J_{kj}\left({\tilde {\bm r}}({\bm y}),\frac{\tau}{2}\right) 
\ B_j({\bf y},(n-1)\tau) e^{-i{\bf K}(n\tau) \cdot {\bf x}} \ 
d^3x d^3y \nonumber \\[2ex]
&\;=\;& \int \int \int \mathcal G^{\eta}_{ik}
\left({\bf x},{\tilde {\bm r}},\frac{\tau}{2}\right) \ 
J_{kj}\left({\tilde {\bm r}}({\bm y}),\frac{\tau}{2}\right) 
\ {\hat B}_j({\bf l},(n-1)\tau) e^{i{\bf l} \cdot {\bm y}} 
e^{-i{\bf K}(n\tau) \cdot {\bf x}} \ \frac{d^3l}{(2\pi)^3} d^3x d^3y,
\end{eqnarray}
where we have also expressed the initial field at time $(n-1)\tau$ in terms
of its Fourier transform. To do the integral over ${\bf x}$, 
we use the identity $\gamma_{jm}(t)\gamma_{mp}(-t) = \delta_{jp}$ to write 
\begin{eqnarray}
\exp{(-i{\bf K}\cdot {\bf x})} &\;=\;& 
\exp{\left( -i K_j\gamma_{jm}(\tau /2) 
[\gamma_{mp}(-\tau /2) x_p- {\tilde r}_m]
-i K_j \gamma_{jm}(\tau /2){\tilde r}_m]\right)}\nonumber\\[2ex]
&\;=\;& \exp{\left[ -i K_j\gamma_{jm}(\tau /2)\left(\tilde{x}_m + \tilde{r}_m\right) \right]} 
\end{eqnarray}
and change the integration variable from 
$x_m$ to $ \tilde {x}_m= \gamma_{mp}(-\tau /2)x_p-{\tilde r}_m $. 
Since $\text{det}|{\bm \gamma}| = 1$, we can write $d^3x = d^3 \tilde x$. 
The integration over $\tilde{\bf x}$ now becomes a Fourier 
transform of the sheared
resistive Greens function \cite{shearGF} and we get
\begin{eqnarray}
\int \mathcal G^{\eta}_{ik}
\left({\bf x},{\tilde {\bm r}},\frac{\tau}{2}\right) e^{-i{\bf K} \cdot {\bf x}}d^3x &\;=\;& \int \frac{\gamma_{ik}(\tau /2)}{(4 \pi \eta \tau)^{3/2}} \left( \text{det}|{\bf D}|\right)^{1/2}\text{exp}\left(\frac{-\tilde x_n D_{nm} \tilde x_m }{4\eta \tau} \right)e^{ -i{\bf K} {\bm \gamma}_{\tau /2} \cdot \tilde {\bf x} } e^{-i{\bf K} \cdot \left[ {\bm \gamma}_{\tau /2}{\tilde {\bm r}}\right]} d^3 \tilde x  \nonumber \\[2ex]
&\;=\;& \gamma_{ik}\left(\tau/2\right) \text{exp}\left\{- \eta \tau  K_n \gamma_{nm}(\tau/2) D^{-1}_{mp} \gamma_{jp}(\tau/2) K_j \right \} e^{-i{\bf K} \cdot \left[ {\bm \gamma}_{\tau/2}{\tilde {\bm r}}\right]} \nonumber \\[2ex]
&\;=\;& \mathcal G_{ik}(\tilde {\bf k})e^{-i \tilde {\bf k} \cdot {\tilde {\bm r}}}
\end{eqnarray}
where $\tilde {k}_m = {K}_j{\gamma}_{jm}(\tau /2) = {k}_j  \gamma_{jm}(-\tau/2)$. Then we have 
\begin{equation}
{\hat B}_i({\bf k}, n\tau) = 
\gamma_{ik}\left(\tau/2\right) \text{exp}\left(- \eta \tau {\tilde k}_p 
D_{pr}^{-1} {\tilde k}_r \right) 
\int \int  J_{kj}\left({\tilde {\bm r}}({\bm y}), \frac{\tau}{2}\right) \ {\hat B}_j({\bf l},(n-1)\tau) e^{i{\bf l} \cdot {\bm y}} 
e^{-i{\bf \tilde k} \cdot {\tilde {\bm r}}} 
 \frac{d^3l}{(2\pi)^3} d^3y\,.
\label{Bfieldk}
\end{equation}

Taking the ensemble average of Eq.(\ref{Bfieldk}), we get for the mean field evolution,
\begin{equation}
{\hat \meanB}_i({\bf k},n\tau) =
\gamma_{ik}\left(\tau/2\right) \text{exp}\left(- \eta \tau {\tilde k}_p 
D_{pr}^{-1} {\tilde k}_r \right) 
\int \left[ \int \overline{e^{-i \tilde {\bf k} \cdot {\tilde {\bm r}}} 
J_{kj}\left({\tilde {\bm r}}({\bm y}), \frac{\tau}{2}\right)} e^{i{\bf l} \cdot y} d^3y \right ] {\hat \meanB}_j({\bf l},(n-1)\tau)
\frac{d^3l}{(2\pi)^3}  
\end{equation}
Here we have assumed as before that the velocity field during the time interval 
$(n\tau,(n-1)\tau)$ and the initial magnetic field at time $(n-1)\tau$ are
statistically independent. Note the dependence on the stochastic 
parameters comes only through ${\tilde {\bm r}}({\bf y})$.
As $\tilde  k_m =   k_j \gamma_{jm}(-\tau/2)$ and  
$\tilde r_m =   \gamma_{mj}(\tau/2) r_j$, we have 
$\tilde {\bf k}\cdot\tilde {\bf r} =  {\bf k}\cdot{\bf r}$  and  
the quantity in the square bracket can be written as
\begin{equation}
\int \overline{e^{-i \tilde {\bf k} \cdot {\tilde {\bm r}}} 
J_{kj}\left({\tilde {\bm r}}({\bm y}), \frac{\tau}{2}\right)} e^{i{\bf l} \cdot y} d^3y  =  \int \overline{e^{-i {\bf k} \cdot ({\bm r}-{\bm y})} 
J_{kj}\left({\tilde {\bm r}}({\bm y}), \frac{\tau}{2}\right)}  e^{-i{\bf k} \cdot {\bf y}} e^{i{\bf l} \cdot {\bf y}}d^3y 
=  \overline{e^{-i {\bf k} \cdot ({\bm r}-{\bm y})} J_{kj}\left({\tilde {\bm r}}({\bm y}), \frac{\tau}{2}\right)} 
(2\pi)^3\delta({\bf l}-{\bf k}).
\end{equation}
In the last step we have used the fact that the averaged quantity is independent of $y_i\,$, as can be easily seen by doing the averaging first over 
the random phase $\psi$ of the turbulent field; this also follows from the statistical homogeneity of the turbulence. The $y_i$ independence is explicitly shown by calculation below. We then have
for the evolution of the mean field,
\begin{equation}
{\hat \meanB}_i({\bf k},n\tau)= G_{ij}({\bf k}){\hat\meanB}_j({\bf k},(n-1)\tau) 
\label{meanb}
\end{equation}
where the response tensor $G_{ij}({\bf k})$ is now
\begin{eqnarray}
G_{ij}({\bf k}) &=& \gamma_{il}(\tau )\overline{\left[\delta_{lj} + {\tilde a}_{l}q_{j} \cos({\bf q} \cdot {\bf y} + \psi)- {\tilde c}_{l}q_{j} \sin({\bf q} \cdot {\bf y} + \psi)\right] e^{-i{\bf k}\cdot ({\bm r} - \textbf{y})}} \exp\left(- \eta \tau  k_p M_{pr}k_r \right)
\label{Gbeforeavgshear}
\end{eqnarray}
where $M_{ij} = \gamma_{im}(-\tau/2) D^{-1}_{mp} \gamma_{jp}(-\tau/2)$ 
and the coefficients ${\tilde a}_{l}$ and ${\tilde c}_{l}$ are shorthand for 
${\tilde a}_{l}(\tau/2, {\bf q})$ and ${\tilde c}_{l}(\tau/2, {\bf q})$, 
given in Eq.(\ref{adefcdef}).
One can see that form of the response tensor in Eq.(\ref{Gbeforeavgshear}) 
with shear is similar to the form of response tensor in Eq.(\ref{Gbeforeavg}) 
without shear and reduces to the latter when $S=0$. Similar to the case
without shear, we see that the effects of resistive dissipation appear only
as a separate exponential term. Since it is small in astrophysical systems of interest, henceforth we set $\eta=0$. 

We now average over the phase $\psi$ of the turbulent velocity field, 
which leads us to the following  expression
\begin{eqnarray}
G_{ij}({\bf k}) &\;=\;& \gamma_{il}(\tau)\, \sigma_{lj}(\textbf{k})
\nonumber \\[2ex] 
\sigma_{lj}(\textbf{k}) &\;=\;&  
\left[\overline{\delta_{lj}\,J_0\left(\sqrt{(\textbf{k} \cdot {\tilde {\bm a}})^2 + (\textbf{k} \cdot {\tilde {\bm c}})^2} \right) \;-\; i\frac{\left({\tilde a}_l q_j\textbf{k} \cdot {\tilde {\bm c}} - {\tilde c}_l q_j\textbf{k} \cdot {\tilde {\bm a}} \right)}{\sqrt{(\textbf{k} \cdot {\tilde {\bm a}})^2 + (\textbf{k} \cdot {\tilde {\bm c}})^2}}J_1\left(\sqrt{(\textbf{k} \cdot {\tilde {\bm a}})^2 + (\textbf{k} \cdot {\tilde {\bm c}})^2} \right)} \right]
\label{Ggen}
\end{eqnarray}
where the overhead bar now refers to the averaging over the directions of ${\bf q}$ and over the randomness of ${\bf A}$. One can reach some conclusions about the decay of the mean magnetic field at this stage of the averaging itself. The mean magnetic field evolves as
\begin{equation}
 \overline{B_i({\bf k},n\tau)} \;=\; \gamma_{il}(\tau ) \sigma_{lj}(\textbf{k})  \overline{B_j({\bf k},0)} 
\end{equation}
The growth or decay of the mean field mode is governed by the product of the two matrices, the shearing matrix ${\bm \gamma}(\tau)$ and the 
shear--turbulence tensor ${\bm \sigma}(\textbf{k})$. It is known that the field grows linearly due to the continuous shearing of the background fluid which causes the $B_1(\textbf{k})$ component of the field to be continuously advected along the ${\bm e_2}$ direction. This is reflected by the presence of the shearing matrix ${\bm \gamma}(\tau)$ in the expression of the response tensor $G({\bf k})$ and is completely natural as well as expected. The turbulent stretching of the field lines due to the transfer of energy from the turbulent pulses of the fluid can too lead to the growth of the mean field and the shear-turbulence tensor ${\bm \sigma}(\textbf{k})$ contains precisely this information through its dependence on the  random parameters. Hence, it is of interest to look at the structure of the ${\bm \sigma}(\textbf{k})$ tensor and calculate its eigenvalues depending on which the mean field grows or decays exponentially.

When helicity is zero, then either ${\tilde {\bm a}}$ or ${\tilde {\bm c}}$ is zero from Eq.(\ref{helicityshear}) and Eq.(\ref{adefcdef}). Then the second term in $\sigma_{lj}(\textbf{k})$ vanishes and we get 
\begin{equation}
\sigma_{lj}(\textbf{k}) \;=\; \delta_{lj} \overline{J_0\left(\textbf{k} \cdot {\tilde {\bm a}} \right)} 
\end{equation}
where we have chosen ${\tilde {\bm c}} = 0$ without any loss of generality. In this case, the eigenvalue of ${\bm \sigma}(\textbf{k})$ is just $\sigma = \overline{J_0\left(\textbf{k} \cdot {\tilde {\bm a}} \right)}$. Note that the maximum value of the Bessel function $J_0(x)$ is unity. Hence after averaging over all the possible values of $\textbf{k} \cdot {\tilde {\bm a}}$ as per the ensemble chosen, we must necessarily obtain  $\overline{J_0\left(\textbf{k} \cdot {\tilde {\bm a}} \right)} < 1 $. This shows that in the absence of helicity 
the mean field modes eventually decay with a decay rate 
$\lambda = (\tau)^{-1}\log\sigma$ (see Eq.(\ref{growthrate})).
Therefore, quite generally, there is no mean field dynamo if the
turbulent velocity is strictly non helical, even in the presence of shear.
 
\subsection{Forced overdamped shearing wave}

We now solve for the form of the $\sigma_{lj}(\textbf{k})$ by taking a particular form of the $\textbf{u}_{\rm turb}$ in Eq.(\ref{usheargen}) obeying the following forced, damped Euler equation:
\begin{eqnarray}
\left( \frac{\partial}{\partial t} \,+\, Sx_1 \frac{\partial}{\partial x_2}  \right)\textbf{u}_{\rm turb} \,+\, Su^1_{\rm turb} {\bm e_2} \,+\, (\textbf{u}_{\rm turb} \cdot \nabla)\textbf{u}_{\rm turb} \;=\; -\nabla p \,-\, \frac{\textbf{u}_{\rm turb}}{\tau_d} \,+\, \textbf{f}
\end{eqnarray}
where $\tau_d$ is a given damping time and $\textbf{f}(\textbf{x},t)$ is the external forcing which is assumed to satisfy $\nabla \cdot \textbf{f} =0$. 
In the approximation $\partial \textbf{u}_{\rm turb} /\partial t \ll \textbf{u}_{\rm turb}/ \tau_d$, the wave is assumed to be overdamped, saturating quickly in time $\tau_d$ to its terminal velocity. In Eq.(\ref{vfinal}) of Appendix~\ref{forcedoverdamped} the following solution is derived:
\begin{eqnarray}
&&\textbf{u}_{\rm turb}({\bf x},t) \;=\; \textbf{A}(t, {\bf q})\,\sin\left[{\bf Q(t)}\cdot {\bf x}+\psi\right] \,+\, \textbf{C}(t, {\bf q})\,\cos\left[{\bf Q(t)}\cdot {\bf x}+\psi\right]\,,\qquad\mbox{where}
\nonumber\\[2ex]
&&A_{1,3} \;=\; a_{1,3} + S\tau_d a_1 \left[\frac{Q_{1,3}Q_2}{Q^2 -S\tau_d Q_1Q_2 } \right]\,,\qquad
A_2 \;=\; a_2 + St a_1 - S\tau_d a_1 \left[\frac{Q_1^2 + Q_3^2}{Q^2 -S\tau_d Q_1Q_2 } \right]\,; \nonumber \\[2ex]
&&C_{1,3} \;=\; h\left\{c_{1,3}+ S\tau_d c_1 \left[\frac{Q_{1,3}Q_2}{Q^2 -S\tau_d Q_1Q_2 } \right] \right\}\,,\qquad
C_2 \;=\; h\left\{c_2 + St c_1 - S\tau_d c_1 \left[\frac{Q_1^2 + Q_3^2}{Q^2 -S\tau_d Q_1Q_2 } \right]\right\}\,,
\end{eqnarray}
where $ Q_j=q_{i}\gamma_{ij}(-t)$, $\textbf{q} \cdot \textbf{a} =0$, $\textbf{c} = \hat{{\bf q}} \times \textbf{a}$ and $h$ such that $-1 \leq h \leq 1$ determines the helicity of the flow. Here the forcing function is related to the constants $\{ a_1, a_2, a_3\}$ and $\{ c_1, c_2, c_3\}$. In the limit of
very strong damping, 
i.e. $\tau_d \to 0$, 
the above expressions simplify and can be written compactly as
\begin{equation}
 A_i \;=\; \gamma_{ij}(t) a_{j}\,,\qquad
C_{i} \;=\; h \gamma_{ij}(t)c_j
\label{adefdamp}
\end{equation}
The turbulent velocity field now becomes 
\begin{equation}
\left[\textbf{u}_{\rm turb}({\bf x},t)\right]_i \;=\; \gamma_{ij}(t) a_{j} \sin\left[{\bf Q(t)}\cdot {\bf x}+\psi\right] \,+\, h \gamma_{ij}(t)c_j \cos\left[{\bf Q(t)}\cdot {\bf x}+\psi\right]
\label{usheardamp}
\end{equation}
Substituting Eq.(\ref{adefdamp}) in Eq.(\ref{adefcdef}), we find that ${\tilde {\bm a}} = {\bf a}\tau $ and ${\tilde {\bm c}} = h{\bf c}\tau\,$; hence the response tensor in Eq.(\ref{Ggen}) becomes
\begin{eqnarray}
G_{ij}({\bf k}) &\;=\;& \gamma_{il}(\tau ) \left[\overline{\delta_{lj}\,J_0\left(\sqrt{(\textbf{k} \cdot  {\bm a} \tau)^2 + (\textbf{k} \cdot {\bm c}h \tau)^2} \right) \;-\; i\frac{\left( a_l q_j h\tau \textbf{k} \cdot {\bm c} - c_l q_j h\tau \textbf{k} \cdot {\bm a} \right)}{\sqrt{(\textbf{k} \cdot {\bm a})^2 + (\textbf{k} \cdot {\bm c}h)^2}}J_1\left(\sqrt{(\textbf{k} \cdot {\bm a}\tau)^2 + (\textbf{k} \cdot {\bm c}h\tau)^2} \right)} \right] \nonumber \\
\end{eqnarray}
As before, let the angle between ${\bf k}$ and ${\bf q}$
be $\theta$; we will treat this as a colatitude and denote the azimuthal angle
of ${\bf q}$ by $\tilde{\phi}$. Let the component of ${\bf k}$ perpendicular to ${\bf q}$ make an angle $\phi$ with ${\bf a}$. Then, on averaging over
the phase $\psi$ we can write
\begin{eqnarray}
G_{ij}({\bf k}) &=& \gamma_{il}(\tau ) \left\{ \overline{\delta_{lj}\,J_{0}(\tau ak \chi \sin\theta) \;-\; \frac{ihaq\tau}{\chi k \sin\theta} \left[\epsilon_{lmn}k_{m}\hat{q}_{n}\hat{q}_{j} \right]J_{1}(\tau ak\chi \sin\theta) } \right \}
\end{eqnarray}
where $\chi = (\cos^2\phi + h^2 \sin^2 \phi)^{1/2}$ and the overhead bar 
denotes ensemble averages over the remaining random variables $\theta$, $\tilde{\phi}$ and $\phi$. Comparing with Eq.(\ref{Ghel}), we see that the form of the response tensor is identical to its form when shear is absent with the only difference being an overall $\gamma_{il}(\tau )$ factor. Since the ensemble average is done at one arbitrary instant, the final form of $G_{ij}({\bf k})$ is identical to Eq.(\ref{Gfinal}) with the extra $\gamma_{ij}$ factor. Hence, we have
\begin{eqnarray}
G_{ij}({\bf k}) &\;=\;& \gamma_{il}(\tau )\,\sigma_{lj}(k)\nonumber\\[2ex]
\sigma_{lj}(k) &\;=\;& \delta_{lj}\ g_{0}(\tau ak, h) \;+\; \frac{ih\tau aq\epsilon_{ljm}k_{m}}{2k} g_{1}(\tau ak, h)\nonumber\\[2ex]
g_{0}(s, h) &\;=\;& \overline{\frac{\sin(s\chi)}{s\chi}}\,;\qquad 
 g_{1}(s, h) \;=\; \overline{\frac{1}{\chi}\left( \frac{\sin(s\chi)}{(s\chi)^2}
-\frac{\cos(s\chi)}{s\chi} \right)}
\label{Gfinalshear}
\end{eqnarray}
and now the overhead bars denote ensembles average over the random variable $\phi$ (for maximally helical flow with $h = \pm 1$, so $\chi = 1$, and the response tensor becomes independent of the random variable $\phi$). 
The shear turbulence tensor ${\bm \sigma}$ has the two non--trivial
eigenvectors $(-i,1,0)^{\rm T}$ and $(i,1,0)^{\rm T}$, with the 
corresponding eigenvalues, $\sigma_+$ and $\sigma_-$ given by
\begin{equation}
\sigma_{\pm}=g_{0}(\tau ak, h) \;\mp\; \frac{\tau aqh}{2}g_{1}(\tau ak, h)
\label{Gfinalshear2}
\end{equation} 
For zero helicity, the second term vanishes and dynamo action is absent, as was shown in the general case in the previous section.
Moreover, even if there were mirror-symmetric fluctuations in $h$,
this would not lead to a dynamo. This is because $g_0(s,h)$ and 
$g_1(s,h)$ are even in $h$, while the co-efficient of the second
term of the response tensor in Eq.~(\ref{Gfinalshear}) is linear 
(and hence odd) in $h$.
Thus on averaging the response tensor over any 
symmetric PDF of $h$ with zero mean
only the first term of $G_{ij}$ survives and there is no dynamo.
This conclusion is similar to that obtained by GB for the case
without shear.

\section{$\alpha^2\Omega$ dynamo}

Let us look at the expression in Eq.(\ref{Gfinalshear}) of the response tensor in the case of small correlation times, when $ak\tau \ll 1$. Then to quadratic order in $\tau$, we get
\begin{eqnarray}
G_{ij}({\bf k}) &\;=\;& 
\delta_{ij}\left[1 \,-\, \frac{(1 + h^2)}{12}(\tau ak)^2\right] \;+\; 
\delta_{i2}\delta_{j1}S \tau \;+\; i\,\frac{ha^2\tau^2 q}{6}\epsilon_{ijm}k_m
\nonumber\\[2ex]
&\;=\;& \delta_{ij}(1-\eta_t \tau k^2) \;+\; \delta_{i2}\delta_{j1}S \tau \;-\;
i\tau \alpha\epsilon_{ijm}k_m  
\end{eqnarray}
where $\alpha = -(1/3)\left[\overline{\textbf{u}_{\rm turb}\cdot (\nabla \times \textbf{u}_{\rm turb})}\right](\tau/2)$ and $\eta_t = (1/3)\left[\overline{\textbf{u}\cdot \textbf{u}}\right](\tau/2)$ are the turbulent transport coefficients. The mean field evolution equation Eq.(\ref{meanb}) then becomes
\begin{equation}
 \left[ {\begin{array}{c}
 B_{\tau 1}  \\[2ex]
 B_{\tau 2}  \\[2ex]
 B_{\tau 3} \\
 \end{array} } \right]  = \left[ {\begin{array}{ccc}
 1-\eta_t \tau k^2 \;&\; -i\alpha \tau k_3  \;&\; i\alpha \tau k_2  \\[2ex]
 i\alpha \tau k_3 + S\tau \;&\; 1-\eta_t \tau k^2 \;&\; -i\alpha \tau k_1  
 \\[2ex]
 -i\alpha \tau k_2 \;&\; i\alpha \tau k_1 \;&\; 1-\eta_t \tau k^2 \\
 \end{array} } \right] \left[ {\begin{array}{c}
 B_{01}  \\[2ex]
 B_{02}  \\[2ex]
 B_{03} \\
 \end{array} } \right] 
\end{equation}
which is the evolution equation for the $\alpha^2\Omega$ dynamo \cite{BS05}. 
We seek solutions of the eigenvalue problem when $k_2 =0$. Of the three eigenvalues, $\sigma_1 = (1-\eta_t \tau k^2)$ is irrelevant, because the 
corresponding eigenvector does not satisfy the solenoidality condition 
$\left(\nabla\cdot{\bf B}\right) = 0\,$. The remaining two eigenvalues are
\begin{equation}
\sigma_{\pm} \;=\; 1 \,-\, \tau\eta_t k^2 \;\pm\; \tau \left(\alpha^2 k^2 -i\alpha k_3 S \right)^{1/2}\,,
\end{equation}
corresponding to the eigenvectors 
\begin{equation}
\left(\alpha k_3\,, \;\pm i\sqrt{\alpha^2 k^2 - i\alpha k_3 S}\,, \;-\alpha k_1\right)^{\rm T}.
\label{eigen}
\end{equation}
The growth rates 
$\lambda_{\pm}$ of these eigenmodes are given by
\begin{equation}
\lambda_{\pm} \;=\; \frac{1}{\tau}\ln(\sigma_{\pm}) \;=\; -\eta_t k^2 \,\pm\, \left(\alpha^2 k^2 -i\alpha k_3 S \right)^{1/2}
\label{growth}
\end{equation}
which are the same as one would get in the case of the $\alpha^2 \Omega$ dynamo \cite{BS05}.

\section{Discussion and Conclusions}\label{conclu}

This paper presents studies of dynamo action in turbulent shear flows
when the turbulence has a non zero correlation time. Our goal is to study
the dynamics of a system which is complex enough to be a useful model, 
yet tractable analytically; the renovating flows discussed earlier by several authors \cite{Ditt,GB92} (and references therein) provide just such a platform. 
Our contribution is to consider random, helical renovating flows in the context of a background linear shear flow. 

We began with a review of the work of Gilbert and Bayly (GB) \cite{GB92} on
random helical renovating flows in the absence of a background shear flow. 
GB considered random flows, each of whose realizations was a plane, 
sinusoidal helical wave. The merit of choosing such simple random ensembles 
is that the trajectories of fluid elements, in the flow caused by 
each wave, can be integrated analytically. Thus the Green's function 
mapping the magnetic field from one time step to another can be obtained, 
and averaged over the underlying 
PDF of the random ensemble of flows. GB give the final result, while skipping
almost all the intermediate steps. We found it useful not only to record 
these missing steps, but lay them out for the reader so that it becomes
easier for us to present our analysis of the more complicated problem of renovating flows with shear. 

We then formulated the problem of renovating flows in the presence of a 
background linear shear flow. Following GB, we considered an 
ensemble of random flows, each of 
whose members is a plane, sinusoidal helical wave. However, unlike in the 
case considered by GB, the wave cannot be time--independent. In fact, each 
of these members must be a \emph{shearing wave}, one whose amplitude and 
wavevector are both time--dependent. Then the trajectories of fluid elements
(in the flow caused by each shearing wave) were determined analytically, 
the Green's function (mapping the magnetic field from one time step to another) 
derived, and averaged over the underlying PDF of the random ensemble of flows,
to obtain a general expression for the (averaged) response tensor. 
We showed that even without fully averaging the response tensor, 
for which one requires the explicit form of the time-dependent, 
shearing wave amplitude, it is still possible to prove a 
general result: that there is no dynamo action when the shearing 
waves are \textit{strictly} non helical.

We then considered a particular model in which the shearing waves were 
generated through external forcing of the linear shear flow. Working in 
the overdamped limit, we derived an explicit form for the response tensor. 
It is interesting to note that this form is closely related to the 
response tensor of GB; specifically, our response tensor is the product 
of the response tensor of GB with the shearing matrix. This was then 
applied to the case of $\alpha^2\Omega$ dynamos in the limit of small 
correlation times, and we     
recovered the standard dispersion relation for the
$\alpha^2\Omega$ dynamo. Thus the growth
of the mean field in sheared helical turbulence is
as expected from quasilinear closures 
usually employed to derive the mean field equations. This 
obtains in spite of the fact that magnetic fluctuations in
such renovating flows are expected
to grow more rapidly (as shown explicitly by GB) than the mean field, 
wherein, one may imagine, that quasilinear closures break down.
Our work therefore provides another illustration that rapidly
growing fluctuations need not destroy the growth of the mean field.

Our result that there is no mean field
dynamo in strictly non helical turbulence, even in the
presence of shear, raises the question as to what causes
such growth in numerical simulations
\cite{shear_sim}. One possibility is the incoherent $\alpha$--shear
dynamo \cite{VB97}. There seems to be evidence 
from such simulations for fluctuations
in $\alpha$ (the first reference in \cite{shear_sim}).
Here the mean fields are defined as averages over two spatial directions,
and fluctuations in $\alpha$ over time are considered as meaningful.
In the present context, we can define the required fluctuations as
due to fluctuations of the parameter $h$ from one renovation time to another,
which is easier to justify as being physically meaningful.
GB themselves considered fluctuations in $h$, but argued that
the PDF of $h$ needs to be skewed for there to be net growth.

Even in the presence of shear, if we averaged the response
tensor say in Eq.~(\ref{Gfinalshear}) over a mirror-symmetric PDF of 
$h$ with zero mean,
the `helical' term would vanish and one
would not have a dynamo. However if  we think of the mean field
as being defined before averaging over $h$, then one could
study its dynamics under such fluctuations.
In the presence of shear, we can see from Eq.~(\ref{growth}), that both
signs of $h$ would cause growth, but the eigenvector 
in Eq.~(\ref{eigen}) would get an extra phase shift. It would be
interesting to work out exactly how such random changes to
the eigenvector alters the efficiency of the dynamo.
One would expect that a coherent $h$ would lead to a more
efficient dynamo rather than a fluctuating $h$.

The work here has focused on the mean field evolution.
The same model can also be used to generalize the
Kazantsev model \cite{Kaz68} for the fluctuation dynamo, to the case
with finite correlation time and including shear.

\section*{Acknowledgements}

SK is supported by a fellowship from the Council of Scientific and Industrial Research (CSIR), India.

\appendix{}


\section{Calculating the averaged response tensor in the absence of shear} \label{averaging}
Below we give the details involved in going from Eq.(\ref{Gbeforeavg}) to Eq.(\ref{Gfinal}). 

\begin{eqnarray}
G_{ij}({\bf k}) \;=\; \overline{\left[\delta_{ij} + a_{i}q_{j}\tau \cos({\bf q} \cdot {\bf y} + \psi) - b_{i}q_{j}h\tau \sin({\bf q}\cdot {\bf y} + \psi)\right] \ e^{-i{\bf k}\cdot ({\bf a}\tau \sin({\bf q}\cdot {\bf y}+\psi)+{\bf b}\tau h \cos({\bf q}\cdot {\bf y}+\psi))}}
\label{gappendix}
\end{eqnarray}
Let the angle between {\bf k} and {\bf q} be $\theta$ . Let the component of {\bf k} perpendicular to {\bf q} make an angle $ \phi $ with {\bf a}.
Now {\bf k} can be written as 

\begin{equation}
{\bf k}=\underbrace{\left[{\bf k}-\frac{{\bf q}({\bf k}\cdot {\bf q})}{q^2}\right]}_{perpendicular \: to \: q} \;+\; \underbrace{\frac{{\bf q}({\bf k}\cdot {\bf q})}{q^2}}_{along \: q}
\end{equation} \\
Since {\bf a} and {\bf b} both lie in the plane perpendicular to {\bf q}, and also since {\bf a} and {\bf b} are perpendicular to each other, 
\begin{eqnarray}
{\bf k}\cdot {\bf a} & = & {\bf k}_{\bot}\cdot {\bf a} = |{\bf k}_{\bot}||{\bf a}|\cos{\phi}=(\text{k}\sin{\theta}) \text{a} \cos{\phi} \nonumber \\
{\bf k}\cdot {\bf b} & = & \text{ka}\sin{\theta} \sin{\phi}
\label{pdef1}
\end{eqnarray}
Now we average $G_{ij}({\bf k})$ term by  term. For the first term, we have
\begin{equation}
I_{1}=\overline{\delta_{ij}  e^{-i{\bf k}\cdot ({\bf a}\tau \sin({\bf q}\cdot {\bf y}+\psi)+{\bf b}\tau h\cos({\bf q}\cdot {\bf y}+\psi))}}
\end{equation}
The argument of the exponential is written as,
\begin{eqnarray}
- i \tau \text{ka}\sin\theta \left[ \cos\phi \sin({\bf q}\cdot {\bf y}+\psi)  +  \text{h}\sin \phi \cos({\bf q}\cdot {\bf y}+\psi)\right]
= - i \tau \text{ka} \chi \sin\theta \sin({\bf q}\cdot {\bf y}+\psi + \alpha) 
\end{eqnarray}
where $\chi = (\cos^2\phi + h^2 \sin^2 \phi)^{1/2}$, $\chi\cos\alpha = 
\cos\phi$ and $\chi\sin\alpha = h\sin\phi\,$. Hence, we have
\begin{equation}
I_{1}=\overline{\delta_{ij}  e^{- i \tau ka \chi \sin\theta \sin({\bf q}\cdot {\bf y}+\psi + \alpha)}}
\end{equation} \\
First we average over $\psi$. Since $\psi$ goes over all the possible phases, on averaging we get
\begin{equation}
I_{1} \;=\; \delta_{ij} \int_{0}^{2\pi} e^{- i (\tau ka \chi \sin\theta) \sin\zeta} \frac{d\zeta}{2\pi} 
\;=\; \delta_{ij}\ J_{0}(\tau ak \chi \sin\theta)
\end{equation}
where we have used the following integral representation of Bessel function of the first kind.
\begin{equation}
\int_{0}^{\pi}e^{i\beta \cos x}\cos(nx)dx=i^n\pi J_{n}(\beta)
\label{bessel}
\end{equation}
We next average over the direction of {\bf q}. We keep {\bf k} fixed (say, along z direction) and vary {\bf q} about {\bf k} over all the solid angles.
\begin{eqnarray}
I_{1}  &\;=\;& \delta_{ij}\ \int_{0}^{\pi} \int_{0}^{2\pi} J_{0}(\tau ak \chi \sin\theta) \frac{\sin\theta d\theta d \tilde \phi}{4\pi} 
\;=\; \ \frac{\delta_{ij}}{2}\ \int_{0}^{\pi} J_{0}(\tau ak \chi \sin\theta) \sin\theta d\theta  \nonumber \\[2ex]
&\;=\;& \ \delta_{ij}\ \int_{0}^{\frac{\pi}{2}} J_{0}(\tau ak \chi \cos\theta) \cos\theta  d\theta  
\;=\; \delta_{ij}\ \frac{\sin(\tau ak \chi)}{\tau ak \chi}  
\end{eqnarray}
Lastly we average over the direction of \textbf{a}. Since \textbf{a} can point in any direction in the plane perpendicular to {\bf q}, we average over the angle $\phi$. We get
\begin{equation}
I_1 \;=\; \delta_{ij}\ g_{0}(\tau ak, h)\,,\quad\mbox{where}\qquad
g_{0}(s, h) \;=\; \overline{\frac{\sin(s\chi)}{s\chi}}
\end{equation}
where the overhead bar now denotes ensemble average over the random variable $\phi$. We proceed to the remaining term of $G_{ij}({\bf k})$ in Eq.(\ref{gappendix}):
\begin{eqnarray}
I_{2}=\overline{\left[a_{i}q_{j}\tau \cos({\bf q}\cdot {\bf y} + \psi) - b_{i}q_{j}h\tau \sin({\bf q}\cdot {\bf y} + \psi)\right] e^{-i{\bf k}\cdot ({\bf a}\tau \sin({\bf q}\cdot {\bf y}+\psi)+{\bf b}\tau h\cos({\bf q}\cdot {\bf y}+\psi))}}
\end{eqnarray}
We proceed in a similar way as above. We define slightly different parameters. The argument of the exponential is now written as,
\begin{eqnarray}
- i \tau ka\sin\theta \left[ \cos\phi \sin({\bf q}\cdot {\bf y}+\psi) + h\sin\phi \cos({\bf q}\cdot {\bf y}+\psi)\right] = i \tau ka \chi \sin\theta \cos({\bf q}\cdot {\bf y}+\psi - \alpha) 
\end{eqnarray}
where $\chi = (\cos^2\phi + h^2 \sin^2 \phi)^{1/2}$, 
$\chi\sin\alpha = -\cos\phi$ and $\chi\cos\alpha = 
-h\sin\phi\,$. Therefore we have
\begin{equation}
I_{2}=\overline{\left[a_{i}q_{j}\tau \cos({\bf q}\cdot {\bf y} + \psi) - b_{i}q_{j}h\tau \sin({\bf q}\cdot {\bf y} + \psi)\right]  e^{i \tau ka \chi \sin\theta \cos({\bf q}\cdot {\bf y}+\psi - \alpha)}}
\end{equation}
First we average over $\psi$. Since $\psi$ goes over all the possible phases, we write
\begin{eqnarray}
I_{2} &=&  \int_{0}^{2\pi} e^{ i (\tau ka \chi \sin\theta) \cos\zeta}\left[a_{i}q_{j}\tau \cos(\zeta + \alpha) - b_{i}q_{j}h\tau \sin(\zeta + \alpha)\right] \frac{d\zeta}{2\pi} \nonumber \\
\end{eqnarray}
Expanding $\cos(\zeta+\alpha)$ and $\sin(\zeta+\alpha)$, and keeping only the the even terms under integration, we get
\begin{eqnarray}
I_{2} &=&  \int_{0}^{2\pi} e^{ i (\tau ka \chi \sin\theta) \cos\zeta}\left[a_{i}q_{j}\tau \cos\alpha \cos\zeta - b_{i}q_{j}h\tau \sin\alpha \cos\zeta\right] \frac{d\zeta}{2\pi} \nonumber \\[2ex]
&=&  i\tau \left[a_{i}q_{j}\cos\alpha - b_{i}q_{j}h\sin\alpha \right]J_{1}(\tau ak\chi \sin\theta) \nonumber \\[2ex]
&=& - \frac{ih\tau q_{j}}{\chi ak\sin\theta} \left[a_{i}({\bf k}\cdot {\bf b}) - b_{i}({\bf k}\cdot {\bf a}) \right]J_{1}(\tau ak\chi \sin\theta)
\end{eqnarray}
where we have used Eq.(\ref{bessel}) to arrive at the second expression and Eq.(\ref{pdef1}) to obtain the last expression. This can be further simplified as
\begin{equation}
q_{j}\left[a_{i}({\bf k}\cdot {\bf b}) - b_{i}({\bf k}\cdot {\bf a}) \right] \;=\; q_{j}\left[{\bf k} \times ({\bf a} \times {\bf b}) \right]_{i} \;=\;
q_{j}\left[{\bf k} \times a^2\hat{{\bf q}} \right]_{i} \;=\;
a^2q\epsilon_{imn}k_{m}\hat{q}_{n}\hat{q}_{j}
\label{rrr}
\end{equation}
We next average over the direction of {\bf q}. We keep {\bf k} fixed along z direction and vary {\bf q} about {\bf k} over all the solid angle
Then we can write $\hat{{\bf q}}=\sin\theta \cos\tilde \phi\ \hat{\textbf{i}}+\sin\theta \sin\tilde \phi\ \hat{\textbf{j}}+\cos\theta \hat{{\bf k}}$
\begin{equation}
I_{2}= - \int_{0}^{\pi} \int_{0}^{2\pi} \frac{ih\tau}{\chi ak\sin\theta} \left[a^2q\epsilon_{imn}k_{m}\hat{q}_{n}\hat{q}_{j} \right]J_{1}(\tau ak\chi \sin\theta) \frac{\sin\theta d\theta d\tilde \phi}{4\pi}
\label{www}
\end{equation}
The $\tilde \phi$ dependence comes only through $\hat{q}_{n}\hat{q}_{j}$, hence we integrate it first over $\tilde \phi$. Note from Eq.(\ref{rrr}) that the z component of $\hat q_{n}$ does not contribute to the integral in Eq.(\ref{www}). Then we get for the x, y components of $\hat q_{n}$, $\hat q_{j}$, $\int_{0}^{2\pi} \hat{q}_{n}\hat{q}_{j}\ \frac{d\tilde \phi}{2\pi}=\frac{\delta_{nj}}{2}\sin^2\theta$.
Thus,
\begin{eqnarray}
-I_{2} &\;=\;&  \int_{0}^{\pi} \frac{ih\tau}{\chi ak\sin\theta} \left[a^2q\epsilon_{i3j}k_{3} \right] \sin^2\theta J_{1}(\tau ak\chi \sin\theta) \frac{\sin\theta d\theta}{4} 
\;=\;  \frac{ih\tau aq\epsilon_{i3j}k_{3}}{\chi 2k} \int_{0}^{\pi} (\sin\theta J_{1}(\tau ak\chi \sin\theta)) \frac{\sin\theta d\theta}{2} \nonumber \\[2ex]
&\;=\;&  \frac{ih\tau aq\epsilon_{i3j}k_{3}}{\chi 2k} \int_{0}^{\pi} \left[-\frac{d}{d(\tau ak \chi)}J_{0}(\tau ak\chi \sin\theta)\right] \frac{\sin\theta d\theta}{2} 
\;=\;  \frac{ih\tau aq\epsilon_{ij3}k_{3}}{\chi 2k} \frac{d}{d(\tau ak \chi)}\int_{0}^{\pi} J_{0}(\tau ak\chi \sin\theta) \frac{\sin\theta d\theta}{2} \nonumber \\[2ex]
&\;=\;&  \frac{ih\tau aq\epsilon_{ij3}k_{3}}{\chi 2k} \frac{d}{d(\tau ak \chi)}\left[\frac{\sin(\tau ak \chi)}{\tau ak \chi}\right] 
\end{eqnarray}
Lastly we average over the direction of \textbf{a}. Since \textbf{a} can point in any direction in the plane perpendicular to {\bf q}, we average over the angle $\phi$. We then get
\begin{equation}
I_{2}= \frac{ih\tau aq\epsilon_{ij3}k_{3}}{2k} g_{1}(\tau ak, h)\,,\quad
\mbox{with}\qquad g_{1}(s, h)=\overline{\frac{1}{\chi}\left(\frac{\sin(s\chi)}{(s\chi)^2}-\frac{\cos(s\chi)}{s\chi}\right)}
\end{equation}
where the overhead bar now denotes ensemble average over the random variable $\phi$.

Combining the results for $I_{1}$ and $I_{2}$, the response tensor is obtained to be
\begin{equation}
G_{ij}({\bf k})= \delta_{ij}\ g_{0}(\tau ak, h) \,+\, \frac{ih\tau aq\epsilon_{ijm}k_{m}}{2k} g_{1}(\tau ak, h)
\end{equation}

\section{Forced overdamped shearing wave} \label{forcedoverdamped}

The forced, damped Euler equation with a background linear shear is (in this Appendix we use $\bfv$ instead of ${\bf u}_{\rm turb}$ for brevity):

\beq
\left(\frac{\partial}{\partial t} \;+\; Sx_1\frac{\partial}{\partial x_2}\right)\bfv \;+\; Sv_1\ey \;+\; \left(\bfv\cdot\nabla\right)\bfv \;=\; -\nabla p \;-\; \frac{\bfv}{\tdamp} \;+\; \bff\,, 
\label{eulereqn}
\eeq
\noindent
where $\tdamp$ is a given damping time, and ${\bf f}(\bfx, t)$ is the external forcing which is assumed to satisfy $\nabla\cdot\bff = 0$. The  pressure, $p(\bfx, t)$, is determined by requiring that Eq.(\ref{eulereqn}) preserve the incompressibility of the flow. We consider external forcing of 
the form  

\beq
\bff \;=\; {\rm Re}\left\{\bfF(t)\,\exp{\left[ i\bfQ(t)\cdot\bfx\right]}\right\}\,;\qquad \bfQ(t)\cdot\bfF(t) \;=\; 0\,,
\label{forcing}
\eeq
\noindent
which excites a single plane shearing wave: 

\begin{eqnarray}
\bfv &\;=\;& {\rm Re}\left\{\bfW(t)\,\exp{\left[ i\bfQ(t)\cdot\bfx\right]}\right\}\nonumber\\[1em]
p &\;=\;& {\rm Re}\left\{P(t)\,\exp{\left[ i\bfQ(t)\cdot\bfx\right]}\right\}
\label{plwform}
\end{eqnarray}
\noindent
Incompressibility (i.e. $\nabla\cdot\bfv = 0$) requires that 

\beq
\bfQ\cdot\bfW \;=\; 0\,, 
\label{incompress}
\eeq
\noindent
which makes the nonlinear term, $\left(\bfv\cdot\nabla\right)\bfv$, vanish because

\beq
\left(\bfW\cdot\nabla\right)\,\exp{\left[\pm i\bfQ(t)\cdot\bfx\right]} \;=\; 
\pm\left( i\bfQ\cdot\bfW\right)\,\exp{\left[ i\bfQ(t)\cdot\bfx\right]} \;=\; 0\,.
\label{nlvanish}
\eeq
\noindent
When Eqs.(\ref{plwform}) are substituted in Eq.(\ref{eulereqn}), we obtain

\beq
\frac{d\bfW}{d t} \;+\;  i\bfW\left(\bfx\cdot\frac{d\bfQ}{d t} + Sx_1Q_2\right) \;+\; SW_1\ey \;=\; - i\bfQ P \;-\; \frac{\bfW}{\tdamp} \;+\; \bfF\,.
\label{reduce}
\eeq
\noindent
Requiring that the $\bfx$--dependent terms inside the parentheses vanish implies that $\bfQ(t)$ must be of the form, 

\beq
Q_1 \;=\; q_1 - St q_2\,;\qquad Q_2 \;=\; q_2\,;\qquad Q_3 \;=\; q_3\,,
\label{Qsoln}
\eeq
\noindent
where $\bfq = (q_1, q_2, q_3)$ is a constant wavevector. Then $\bfW(t)$ satisfies

\beq
\frac{d\bfW}{d t} \;+\; SW_1\ey \;=\; - i\bfQ P
 \;-\; \frac{\bfW}{\tdamp} \;+\; \bfF\,. 
\label{Aeqn}
\eeq
\noindent
where $Q^2 = \bfQ\cdot\bfQ = (q_1 - St q_2)^2 + q_2^2 + q_3^2\,$. We consider the overdamped case when $\left\vert d\bfW/d t\right\vert \,\ll\, \left\vert\bfW/\tdamp\right\vert$, so we drop the time derivative term on the left side of Eq.(\ref{Aeqn}). $P$ can now be eliminated by taking the dot product of Eq.(\ref{Aeqn}) with $\bfQ$ and using $\bfQ\cdot\bfW = 0$. Then $\bfW(t)$ satisfies

\beq
SW_1\ey \;=\; SW_1\left(\frac{Q_2\bfQ}{Q^2}\right) \;-\; \frac{\bfW}{\tdamp} \;+\; \bfF\,. 
\label{Aodeqn}
\eeq
\noindent
The solution is
\begin{eqnarray}
W_1 &\;=\;& \tdamp F_1 \;+\; S\tdamp\left[\frac{Q_1Q_2}{Q^2 - S\tdamp Q_1Q_2}\right]\tdamp F_1\nonumber\\[1em]
W_2 &\;=\;& \tdamp F_2 \;-\; S\tdamp\left[\frac{Q_1^2 + Q_3^2}{Q^2 - S\tdamp Q_1Q_2}\right]\tdamp F_1\nonumber\\[1em]
W_2 &\;=\;& \tdamp F_3 \;+\; S\tdamp\left[\frac{Q_2Q_3}{Q^2 - S\tdamp Q_1Q_2}\right]\tdamp F_1
\label{Asoln}
\end{eqnarray}
\noindent
Using $\bfQ\cdot\bfF = 0$, it can be verified that $\bfQ\cdot\bfW = 0$. The above solution for $\bfW$ is valid for quite arbitrary forms of the forcing. Now we make a specific choice for $\bfF(t )$: 
\beq
\tdamp F_1 \;=\; G_1\,;\qquad
\tdamp F_2 \;=\; G_2 + St  G_1\,;\qquad \tdamp F_3 \;=\; G_3\, 
\label{Fchoice}
\eeq
\noindent
where $\bfG = (G_1, G_2, G_3)$ is a constant complex vector that is orthogonal to $\bfq$ (i.e.  $\bfq\cdot\bfG = 0\,$). Then the dependence of $\bfW$ on the time $t $ is given in explicit form as
\begin{eqnarray}
W_1 &\;=\;& G_1 \;+\; S\tdamp\left[\frac{Q_1Q_2}{Q^2 - S\tdamp Q_1Q_2}\right]G_1\nonumber\\[1em]
W_2 &\;=\;& G_2 + St  G_1 \;-\; S\tdamp\left[\frac{Q_1^2 + Q_3^2}{Q^2 - S\tdamp Q_1Q_2}\right]G_1\nonumber\\[1em]
W_2 &\;=\;& G_3 \;+\; S\tdamp\left[\frac{Q_2Q_3}{Q^2 - S\tdamp Q_1Q_2}\right]G_1
\label{Aexp}
\end{eqnarray}
\noindent
We now write the velocity field in explicit real form, using arguments familiar from the discussion  of the polarization of monochromatic plane electromagnetic waves. $\bfG$ is a complex vector. If its square, $G^2 = \bfG\cdot\bfG$, has argument equal to $2\psi$, then we may write, $\bfG = \bfE\,\exp{[ i\psi]}$ with $\bfq\cdot\bfE = 0\,$, where $\bfE$ is a complex vector whose square, $E^2 = \bfE\cdot\bfE$, is a real quantity. We now express $\bfE$ in explicit form as, $\bfE = (h\bfc -  i\bfa)$ with $\bfq\cdot\bfa = 0$ and $\bfq\cdot\bfc = 0\,$, where $\bfc$ and $\bfa$ are real vectors orthogonal to $\bfq$, and $h$ is a real number; we can choose $|\bfc| = |\bfa|$ and $-1\leq h\leq 1\,$. Since $E^2 = \left(h^2c^2 - a^2 -2 i h\,\bfc\cdot\bfa\right)$ has been chosen to be a real quantity, we must have $\bfc\cdot\bfa = 0$. In other words, $\bfa$ and $\bfc$ are mutually orthogonal vectors lying in the plane perpendicular to $\bfq$. Then the velocity field of the sheared plane wave of Eq.(\ref{plwform}) is given by, 

\begin{eqnarray}
&&\textbf{v}({\bf x},t) \;=\; \textbf{A}(t, {\bf q})\,\sin\left[{\bf Q(t)}\cdot {\bf x}+\psi\right] \,+\, \textbf{C}(t, {\bf q})\,\cos\left[{\bf Q(t)}\cdot {\bf x}+\psi\right]\,,\qquad\mbox{where}
\nonumber\\[2ex]
&&A_{1,3} \;=\; a_{1,3} + S\tau_d a_1 \left[\frac{Q_{1,3}Q_2}{Q^2 -S\tau_d Q_1Q_2 } \right]\,,\qquad
A_2 \;=\; a_2 + St a_1 - S\tau_d a_1 \left[\frac{Q_1^2 + Q_3^2}{Q^2 -S\tau_d Q_1Q_2 } \right]\,; \nonumber \\[2ex]
&&C_{1,3} \;=\; h\left\{c_{1,3}+ S\tau_d c_1 \left[\frac{Q_{1,3}Q_2}{Q^2 -S\tau_d Q_1Q_2 } \right] \right\}\,,\qquad
C_2 \;=\; h\left\{c_2 + St c_1 - S\tau_d c_1 \left[\frac{Q_1^2 + Q_3^2}{Q^2 -S\tau_d Q_1Q_2 } \right]\right\}\,,\nonumber \\[2ex]
&&\mbox{where}\quad Q_1 \;=\; q_1 - St  q_2\,,\quad Q_2 \;=\; q_2\,,\quad Q_3 \;=\; q_3\,;\qquad
\bfq\cdot\bfa \;=\; 0\,,\quad\bfc \;=\; \hat{\bfq}\times\bfa\,,\quad -1\leq h\leq 1\,.
\label{vfinal}
\end{eqnarray}

\bibliographystyle{unsrtnat}
\bibliography{magnetic}

\end{document}